\documentclass[%
reprint,
superscriptaddress,
nofootinbib,
 amsmath,amssymb,
 aps,
prb,
]{revtex4-2}

\usepackage{graphicx}% Include figure files
\usepackage{dcolumn}% Align table columns on decimal point
\usepackage{bm}% bold math
\usepackage{chemformula}% chemie symbols
\usepackage{textcomp}
\usepackage{xcolor}
\usepackage[normalem]{ulem}
\usepackage{braket}
\usepackage{upgreek}
\usepackage{placeins}
\usepackage[pagebackref=false]{hyperref}
\usepackage{upgreek}

\begin{document}
\title{Ultra-steep slope cryogenic FETs based on bilayer graphene}

\author{E.~Icking}
\affiliation{JARA-FIT and 2nd Institute of Physics, RWTH Aachen University, 52074 Aachen, Germany,~EU}%
\affiliation{Peter Gr\"unberg Institute  (PGI-9), Forschungszentrum J\"ulich, 52425 J\"ulich,~Germany,~EU}

\author{D.~Emmerich}
\affiliation{JARA-FIT and 2nd Institute of Physics, RWTH Aachen University, 52074 Aachen, Germany,~EU}%
\affiliation{Peter Gr\"unberg Institute  (PGI-9), Forschungszentrum J\"ulich, 52425 J\"ulich,~Germany,~EU}

\author{K.~Watanabe}
\affiliation{Research Center for Electronic and Optical Materials, National Institute for Materials Science, 1-1 Namiki, Tsukuba 305-0044, Japan
}
\author{T.~Taniguchi}
\affiliation{ 
Research Center for Materials Nanoarchitectonics, National Institute for Materials Science,  1-1 Namiki, Tsukuba 305-0044, Japan
}
\author{B.~Beschoten}
\affiliation{JARA-FIT and 2nd Institute of Physics, RWTH Aachen University, 52074 Aachen, Germany,~EU}%
\author{M.~C.~Lemme}
\affiliation{Chair of Electronic Devices, RWTH Aachen University, 52074 Aachen, Germany,~EU}
\affiliation{AMO GmbH, 52074 Aachen, Germany,~EU}
\author{J.~Knoch}
\affiliation{IHT, RWTH Aachen University, 52074 Aachen, Germany,~EU}
\author{C.~Stampfer}
\affiliation{JARA-FIT and 2nd Institute of Physics, RWTH Aachen University, 52074 Aachen, Germany,~EU}%
\affiliation{Peter Gr\"unberg Institute  (PGI-9), Forschungszentrum J\"ulich, 52425 J\"ulich,~Germany,~EU}%


\begin{abstract}
Cryogenic field-effect transistors (FETs) offer great potential for a wide range of applications, the most notable example being classical control electronics for quantum information processors.
In the latter context, on-chip FETs with low power consumption are a crucial requirement. This, in turn, requires operating voltages in the millivolt range, which are only achievable in devices with ultra-steep subthreshold slopes. 
However, in conventional cryogenic metal-oxide-semiconductor (MOS)FETs based on bulk material, the experimentally achieved inverse subthreshold slopes saturate around a few mV/dec due to disorder and charged defects at the MOS interface.
FETs based on two-dimensional materials offer a promising alternative.
Here, we show that FETs based on Bernal stacked bilayer graphene encapsulated in hexagonal boron nitride and graphite gates exhibit inverse subthreshold slopes of down to 250\,$\upmu$V/dec at 0.1~K, approaching the Boltzmann limit.
This result indicates an effective suppression of band tailing in van-der-Waals heterostructures without bulk interfaces, leading to superior device performance at cryogenic temperature.
\end{abstract}

\maketitle
Field-effect transistors operable at cryogenic temperatures are an ongoing area of research with potential applications in outer space electronic devices~\cite{Gutierrez2001, chen2006, patterson2006, Bourne2008, han2022feb}, semiconductor-superconducting coupled systems~\cite{Feng2004}, scientific instruments such as infrared sensors~\cite{Zocca2009, Wada2012,han2022feb}, and notably control electronics in quantum computing~\cite{ Homulle2017, Almudever2017, vandersypen2017Sep, Patra2018, Incandela2018, Boter2022Aug}. 
The distinct advantages of operating at cryogenic temperatures include reduced power dissipation, minimized thermal noise, and faster signal transmission~\cite{ Gutierrez2001, Balestra2001, Rajashekara2013}.  
The significance of cryogenic control electronics is especially apparent 
in the context of quantum information processing, where the availability of control electronics in close proximity to the qubits is seen as a necessary condition for operating large quantum processors with thousands of qubits.~\cite{ Charbon2016Feb, vandersypen2017Sep, Galy2018may, Hollmann2018Nov, Guevel2020feb, Boter2022Aug}. 
However, developing cryogenic electronics for quantum computing applications poses significant challenges due to the limited cooling power of dilution refrigerators.
One of the requirements is to reduce the operational voltage range of the FETs into the mV range~\cite{Knoch2023Apr}, which, in turn, requires devices with ultra-steep subthreshold slopes. Temperature broadening effects impose
a lower limit -- the so-called Boltzmann limit -- to the inverse subthreshold slope (SS) given by SS$_\text{BL}$ $= k_\text{B}T/e\cdot\ln(10)$, where $T$ is the operating temperature and $k_\text{B}$ the Boltzmann constant. Thus the inverse SS is expected to decrease from 60\,mV/dec at room temperature to as low as, e.g., 20\,$\upmu$V/dec at 0.1\,K.
However, experiments with  conventional FET devices optimized for low-temperature operation have shown that the inverse SS saturates at considerably higher values in the order of 10\,mV/dec at cryogenic temperature~\cite{Achour2013, Bohuslavskyi2019Mar, Beckers2020jan, Beckers2020dec}.
This saturation originates mainly from static disorder at the metal-oxide-semiconductor (MOS) interface (due to, e.g., surface roughness, charged
defects, etc.)~\cite{Bohuslavskyi2019Mar, Beckers2020jan, Beckers2020dec, kamgar1982, Beckers2020jan, ghibaudo2020aug}. This contributes to the formation of a finite density of states (DOS) near the band edges, which decays exponentially into the band gap~\cite{hill1978}. This so-called band-tailing leads to deteriorated off-state behavior and limits the achievable SS. This effect is further enhanced by dopants, which could either freeze out or become partially ionized~\cite{Beckers2019, Knoch2023Apr}.
Interface engineering can improve the MOS interface~\cite{Richtstein2022Oct}, but in MOSFETs based on bulk materials, inherent disorder at the interfaces and charged defects within bulk dielectrics cannot be fully eliminated. 
%%%%%%%%%%%%%%%%%%%%%%%%% Figure 1
\begin{figure*}[hbt]
    \centering
    \includegraphics[width=0.75\linewidth]{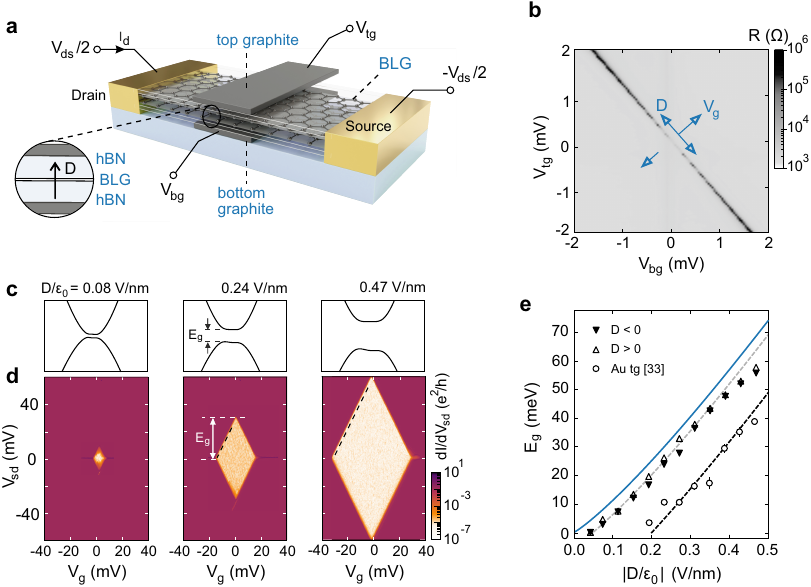}
    \caption{\textbf{a} Schematic illustration of a bilayer graphene-based FET.
    In the active area of the device, the hBN-BLG-hBN heterostructure (see inset) is sandwiched between 
 a top and bottom graphite gate. 
 These gates allow for an independent tuning of the displacement field $D$ and the effective gate voltage $V_\text{g}$.
 The drain-source voltage $V_\text{ds}$ is applied symmetrically in all our measurements. \textbf{b} Resistance ($R=V_\text{ds}/I_\text{d}$) of the BLG as a function of $V_\text{bg}$ and $V_\text{tg}$ at $T = 1.6$~K and $V_\text{ds}=1$~mV.
 The blue arrows indicate the directions of increasing displacement field $D$ and $V_\text{g}$.
 \textbf{c} Calculated band structure of BLG around one of the band minima for different displacement fields (see labels).
 \textbf{d}
 Differential conductance $dI/dV_\text{ds}$ as a function of $V_\text{ds}$ and $V_\text{g}$ (at $T = 0.1$~K) for different displacement fields (see labels in c).
 The band gap $E_\text{g}$ can be extracted from the extension of the diamond along the $V_\text{ds}$ axis (see label). 
 \textbf{e} Extracted $E_\text{g}$ as a function of $|D/\varepsilon_0|$. The experimental data are in good agreement with theory calculated according to Ref.~\cite{McCann_2013} using $\varepsilon_\text{BLG}=1$ (blue line) including an offset of 5\,meV (grey dashed line). Note, that for the same displacement field, the achieved band gap is almost 20\,meV higher compared to state-of-the-art BLG devices with gold top gates (open circles taken from Ref.~\cite{Icking2022Oct}).}
    \label{fig:f1}
\end{figure*}
%%%%%%%%%%%%%%%%%%%%%%%%%

%%%%%%%%%%%%%%%%%%%%%%%%% Figure 2
\begin{figure*}[hbt]
    \centering
    \includegraphics[width=0.82\linewidth]{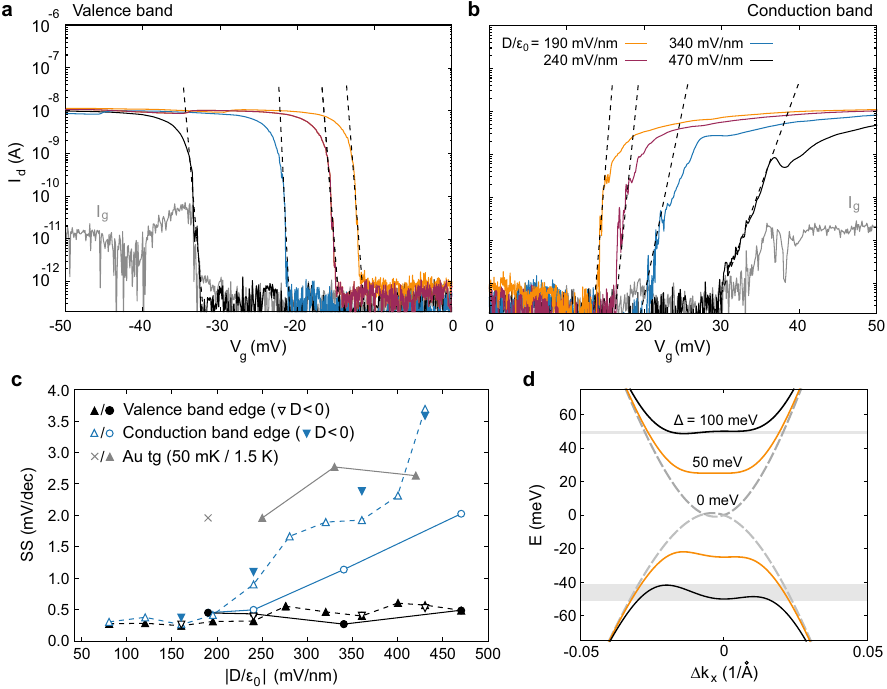}
    \caption{\textbf{a,b} Drain current as a function of $V_\text{g}$ for four different displacement fields (see different colors and labels in panel b) near the valence band edge (panel a) and the conduction band edge (panel b). The gate leakage current is shown as the gray trace exemplarily for $D/\varepsilon_0=470$\,mV/nm (see also Supp. Mat. Fig.~S3).
    Measurements were taken at $V_\text{ds}=0.1$\,mV and $T=0.1$\,K. \textbf{c} Extracted minimal subthreshold slope as a function of the displacement field for both, the valence (black) and conduction (blue) band edges. The black-filled circles and blue circles correspond to data directly extracted from the measurements shown in panels a and b (see black dashed lines), respectively. 
    The upwards-pointing triangles are extracted from similar measurements at slightly higher $V_\text{ds}\approx 0.5$\,mV. 
    Both measurements result in values around 0.3\,mV/dec at the valence band edge. At the conduction band edge the SS$_\text{min}$ values show an increase with increasing $D$. 
    Downwards-pointing triangles denote SS extracted for negative displacement fields.
    The gray symbols represent the SS extracted from two devices with a gold top gate at the valence band edge (cross: 1st device measured at 50\,mK, gray upwards-pointing triangles: 2nd device measured at 1.5\,K). \textbf{d} Calculated band structure for different onsite potential differences $\Delta$ between the BLG layers. $\Delta k_x$ represents the momentum relative to the K and K' points. Due to trigonal warping effects~\cite{Varlet2014}, the bands show an asymmetric deformation if a band gap is present. With increasing onsite potential difference, the asymmetry of the deformation increases, indicating a possible origin of the asymmetry in subthreshold slope values.}
    \label{fig:f2}
\end{figure*}
%
%\newline 
FETs based entirely on van der Waals (vdW) materials are a promising alternative because these materials offer atomically clean interfaces, as there are no dangling bonds in the vertical direction. Particularly promising for cryogenic applications are vdW-heterostructures based on Bernal stacked bilayer graphene (BLG)~\cite{Knoch2023Dec}. Indeed, it has been shown that by encapsulating BLG into hexagonal boron nitride (hBN) and by placing it on graphite (Gr), it is possible to open a tunable, ultraclean, and spatially homogeneous band gap in BLG by applying an out-of-plane electric displacement field.~\cite{McCann2006Mar, McCann_2013, Jung2014Jan}. Such BLG-based heterostructures can be seen as an electrostatically tunable semiconductor~\cite{Li2016, Overweg2018Dec, Icking2022Oct}.
The high device quality allowed the realisation of BLG-based quantum point contacts~\cite{Overweg2018Dec, Banszerus2020May} and quantum dot devices~\cite{Eich2018Aug, Banszerus2018Aug,banszerus2023may}. Further incorporating graphite top gates (tg) instead of state-of-the-art gold top gates in the BLG heterostructures promises a further reduction of disorder as recent publications reported magnetic and even superconducting phases hosted in the valence and conduction bands of BLG~\cite{Zhou2022Jan,seiler2022Aug,barrera2022Jul}.\newline
In this work, we demonstrate the enhanced device quality of dual graphite-gated BLG,
evident in ultra-clean band gaps and ultra-small inverse subthreshold slopes, establishing vdW-material-based heterostructures as an ideal platform for cryogenic FETs. We use finite bias spectroscopy to show that the band gap tunability is enhanced in pure vdW BLG heterostructures with almost no residual disorder. By extracting the inverse subthreshold slopes, we obtain values as low as 250\,$\upmu$V/dec at $T=0.1$\,K, which is only an order of magnitude larger than the Boltzmann limit of 20\,$\upmu$V/dec at this temperature. These results demonstrate the effective suppression of band tailing, leading to superior cryogenic device behavior of FETs based on vdW materials compared to conventional FETs.\\
%%%%%%%%%%%%%%%%%%%%%%%%% Figure 3
\begin{figure*}[hbt]
    \centering
    \includegraphics[width=0.7\linewidth]{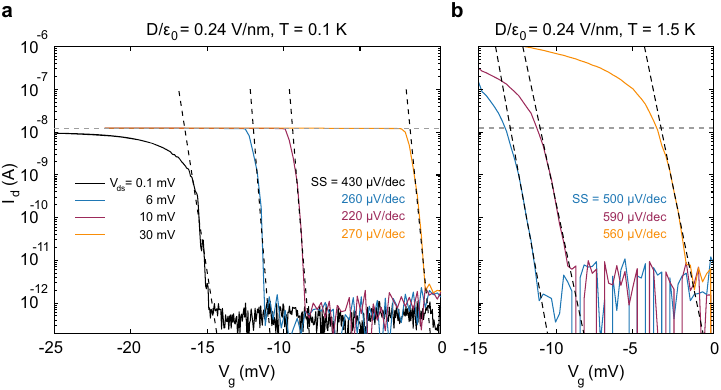}
    \caption{\textbf{a,b} Drain current as a function of $V_\text{g}$ at the valence band edge for different applied drain-source voltages $V_\text{ds}$ at a fixed displacement field $D/ \varepsilon_0 \approx0.24$~V/nm. The data shown in panel \textbf{a} were taken in a dilution refrigerator at $T=0.1$\,K, while those presented in panel \textbf{b} were taken in a pumped $^4$He cryostat at $T=1.5$\,K. The first setup limits the on-current to roughly $10^{-8}$\,A. The second system allows higher on-currents of 1\,$\upmu$A. However, we observe a higher noise level resulting in a slightly increased off-current.
    }
    \label{fig:f3}
\end{figure*}
%%%%%%%%%%%%%%%%%%%%%%%%%
\newline
The studied devices are fabricated by a standard dry
van-der-Waals transfer technique~\cite{Wang2013Nov, Purdie2018}. The process involves the sequential stacking of hBN, graphite and BLG flakes produced by mechanical exfoliation~\cite{Novoselov2004}.
First, a large hBN flake is selected to completely cover the top graphite gate, which is picked up in the second step. The (top) graphite gate is encapsulated in another hBN flake which acts as the top gate dielectric. We then pick up the BLG, a third hBN flake (bottom gate dielectric), and the bottom graphite gate and transfer the vdW heterostructure to a \ch{Si}$^{++}$/\ch{SiO2} substrate. 
The exact thicknesses of the used hBN dielectric layers (mainly $\approx 20$~nm) can be found in the Supp. Mat. Tab.~S1.
Complete encapsulation
of the BLG in hBN is essential to prevent degradation
and short circuits to the graphite gates. 
One-dimensional side contacts are then fabricated using electron-beam lithography,
CF$_4$-based reactive ion etching and metal evaporation followed
by lift-off~\cite{Wang2013Nov}. A schematic of the final device, including the gating and contacting scheme, is shown in Fig.~\ref{fig:f1}a (an optical image can be found in the Supp. Mat. Fig.~S1). 
If not stated otherwise, all measurements were performed at $T=0.1$\,K in a dilution refrigerator with a two-terminal configuration, where we applied the drain-source voltage symmetrically (for more information on the measurement setup, see Ref.~\cite{Icking2022Oct}).\newline
%
%
%
%~
%
As a first electrical characterization, we measure the drain current $I_\text{d}$ as a function of top and bottom gate voltage by applying a small drain-source voltage $V_\text{ds}=100$\,$\upmu$V. 
Fig.~\ref{fig:f1}b shows the resulting map of the BLG resistance $R=V_\text{ds}/I_\text{d}$. Here, we observe a diagonal feature of increased resistance with a slope $\beta=1.22$, which gives us directly the relative gate lever arm $\beta=\alpha_\mathrm{bg}/\alpha_\mathrm{tg}$, where $\alpha_\mathrm{bg}$ and $\alpha_\mathrm{tg}$ denote the gate lever-arms of the top and bottom gate and can be extracted from quantum Hall measurements~\cite{Zhao2010, SonntagMay2018, Schmitz2020} (for more information, see Supp. Mat.). The increasing width of the region of maximum resistance with increasing gate voltages is direct evidence for the formation and tuning of the BLG band gap with increasing out-of-plane displacement field $D$ (see also band structure calculations in Fig.~\ref{fig:f1}c).
The displacement field in the dual-gated BLG-based vdW heterostructure is given by $D=e\alpha_\mathrm{tg}\left[\beta\left(V_\mathrm{bg}-V^0_\mathrm{bg}\right)-\left(V_\mathrm{tg}-V^0_\mathrm{tg}\right)\right]/2$, 
and the effective gate voltage is given by $V_\mathrm{g} = \left[\beta\left( V_{bg} - V_\mathrm{bg}^0\right) + \left( V_\mathrm{tg} - V_\mathrm{tg}^0 \right)\right]/(1 + \beta)$, which tunes the electrochemical potential in the band gap of the BLG, $\mu \approx e V_\text{g}$~\cite{Icking2022Oct}. Here, $\varepsilon_0$ is the vacuum permittivity, and the parameters $V_\mathrm{tg}^0$ and $V_\mathrm{bg}^0$ account for the offsets of the charge neutrality point from $V_\mathrm{tg}=V_\mathrm{bg}=0$. \\
~\newline
To study the band gap opening in our devices as a function of the displacement field $D$, we perform finite bias spectroscopy measurements and investigate the differential conductance $dI/dV_\text{ds}$ as a function of the effective gating potential $V_\text{g}$ and the applied drain-source voltage $V_\text{ds}$ for different fixed displacement fields $D$, see Fig.~\ref{fig:f1}d.
A distinct diamond-shaped region of suppressed conductance emerges, which has a high degree of symmetry and sharp edges and scales well with the applied displacement field. The outlines of the diamonds (black dashed lines in Fig.~\ref{fig:f1}d) show a slope of $\approx2$, highlighting that $V_\text{g}$ directly tunes the electrochemical potential $\mu$ within the band gap and indicating that the band gap is as good as free of any trap states~\cite{Icking2022Oct}. 
In the Supp. Mat. we show that the slope of the diamond outlines is indeed constant ($\approx 2$) for all displacement fields $D \gtrsim 0.2$\,V/nm. \\
~\newline
From the extension of the diamonds on the $V_\text{ds}$ axis, we can directly extract the size of the band gap $E_\text{g}$~\cite{Icking2022Oct}, which are shown in Fig.~\ref{fig:f1}e for positive (filled triangles) and negative displacement fields (empty triangles). They agree reasonably well with the theoretical prediction assuming an effective dielectric constant of BLG of $\varepsilon_\text{BLG}=1$ (blue line, for more information, see Supp. Mat.) except for a small offset of 5\,meV (gray dashed line), which might be due to some residual disorder or interaction effects. 
Measurements on a second graphite top-gated device reveal the same behavior (see Supp. Mat. Fig.~S8).\\ 
~\newline
In Fig.~\ref{fig:f1}e we also report the results of measurements performed on a similar BLG device but with the top gate made of gold instead of graphite (see Ref.~\cite{Icking2022Oct}).  It is noteworthy that the extracted band gap for the device with graphite gates is almost 20\,meV higher than that extracted for the device with a gold top gate for the same displacement fields, highlighting the importance of clean vdW-interfaces. 
Furthermore, the observed extracted band gap $E_\text{g}$ persists down to lower displacement fields $D/\varepsilon_0\approx50$\,mV/nm compared to devices with a gold top gate.

~\newline
The high tuning efficiency of the band gap in graphite dual-gated BLG combined with the high symmetry of the diamonds from the bias spectroscopy measurements demonstrates that BLG heterostructures built entirely from vdW materials, including top and bottom gates, outperform BLG devices with non-vdW materials thanks to much cleaner interfaces, allowing them to achieve unprecedented levels of device quality.
~\\
\newline
The finite bias spectroscopy measurements show that the edges of the diamonds are sharply defined, which promises excellent switching efficiency of FETs based on dual graphite-gated BLG when using $V_\text{g}$ as the tuning parameter. To extract the inverse subthreshold slope, we measure the drain current $I_\text{d}$ as a function of $V_\text{g}$ for fixed $D$-field and $V_\text{ds}\approx0.1$\,mV at both band edges, see Figs.~\ref{fig:f2}a and \ref{fig:f2}b.
From the linear fits of the slopes (black dashed lines), we extract the inverse subthreshold slope SS~$=\left(\partial(\log_{10}(I_\text{d})/\partial V_\text{g}\right)^{-1}$. The resulting values for the valence and conduction band are plotted in Fig.~\ref{fig:f2}c.\newline
At the valence band edge, we extract record low values of SS $\approx270$~-~$500$\,$\upmu$V/dec, roughly one order of magnitude above the Boltzmann limit SS$_\text{BL} (\text{0.1\,K})=20$\,$\upmu$V/dec. For comparison, the saturation limit of conventional FETs based on non-vdW materials at $T\approx$ 0.1\,K is in the order of a few mV/dec~\cite{Beckers2020dec}. We repeat similar measurements for slightly higher drain-source voltages $V_\text{ds}\approx0.5$\,mV. The results are also shown in Fig.~\ref{fig:f2}c as upwards-pointing triangles. They agree overall with the values from the measurements at $V_\text{ds}=0.1$\,mV, with inverse subthreshold slopes at the valence band around SS $\approx250$ to $500$\,$\upmu$V/dec. The very low SS value indicates that band tailing is suppressed for devices with only vdW interfaces. This is also supported by the fact that samples with a gold top gate (i.e. an interface between a vdW and a bulk material) show significantly higher SS values for comparable $D$-fields at the valence band edge (see the cross and gray upward-pointing triangles in Fig.~\ref{fig:f2}c). 
\newline
It is remarkable to observe that while the SS extracted at the valence band edge does not show a significant dependency on the applied displacement field $D$, while the values extracted at the conduction band edge show a considerable increase from SS $\approx500$\,$\upmu$V/dec up to SS $\approx2.8$\,mV/dec with increasing $D$.  
This displacement field-dependent asymmetry of the SS values is related to the electron-hole asymmetry of the BLG band structure. In principle, this asymmetry could also be due to a top-bottom asymmetry of (weak) interface disorder in the vdW heterostructure, since transport near the band edges is dominated by orbitals in only one of the two graphene layers. For example, for a positive $D$-field, transport at the conductance (valence) band edge is carried only by the top (bottom) layer of the BLG~\cite{Icking2022Oct}. Changing the $D$-field direction reverses the band-edge to layer assignment. This allows us to experimentally exclude such a possible non-uniformity of the interface disorder, as we observe the same asymmetry in the SS values for the conductance and valence band edge also for negative $D$-fields (see downward pointing triangles in Fig.~\ref{fig:f2}c), in good agreement with the values for positive $D$-fields, thus strongly emphasizing the importance of the asymmetry in the BLG band structure.
In Fig.~\ref{fig:f2}d we show the calculated
band structure as a function of the onsite potential difference between the layers $\Delta (D)$, which can be directly tuned with the applied displacement field $D$ (for more information on the calculations, see Supp. Mat.). With increasing $\Delta (D)$, the bands undergo an increasingly asymmetric deformation due to the trigonal-warping effect~\cite{Varlet2014,varlet2015dec}. As a consequence, the bands change from a hyperbolic shape at low $\Delta (D)$ to an asymmetric Mexican-hat shape for high $\Delta (D)$~\cite{McCann_2013}, see Fig.~\ref{fig:f2}d. With increasing band deformation, parts of the bands close to the $K$ and $K'$ points of the Brillouin zone become flat. Recent studies have shown that these flat bands give rise to a rich phase diagram in BLG, where magnetic and superconducting phases emerge~\cite{Zhou2022Jan, seiler2022Aug, barrera2022Jul}. The emerging phases could act phenomenologically similar to the interface-induced disorder, resulting in effective tail states at the band edges and degradation of the SS. The flat parts of the bands are right at the conduction band edge, but slightly deeper in the valence band: for example, for $\Delta=100$\,meV in Fig.~\ref{fig:f2}d, the local valence band maximum is much more pronounced than the local conduction band minimum (see grey shaded areas). Consequently, the resulting phase diagrams also exhibit an asymmetry similar to our SS values~\cite{barrera2022Jul}, which suggests that the asymmetric band deformation could cause the SS asymmetry in our measurements. 
We observed the same behavior for a second device, although at slightly different $D$-fields (see Supp. Mat. Fig.~S9), most likely due to sample-to-sample variations.
Regardless, we would like to emphasize that this consistent asymmetry is in itself an indicator of the overall low disorder in our devices.
\\
\newline
While our device presents excellent SS values, the measured on-off ratio in Fig.~\ref{fig:f2}a and b is only about 10$^4$ to 10$^5$, which is a direct consequence of the low on-current of about $I_\text{d}\approx10^{-8}$\,A. 
This low current level is partially due to the small size of the device contacts, which are circularly etched vias through the hBN, with a diameter of just 1\,$\upmu$m.
However, it is mainly because the measured current is limited by our measurement setup, which is optimized for low-noise, small-current measurements but also imposes a sharp limit of about $10^{-8}$\,A, see Fig.~\ref{fig:f3}a. 
In a different setup at higher temperatures $T=1.5$\,K, we observe on-currents of up to 1\,$\upmu$A in the very same device for large $V_\text{ds}=30$\,mV, see Fig.~\ref{fig:f3}b, indicating that higher currents are possible also with the contact geometry used.
This is also confirmed by measurements in a second device of similar design, where we measure currents up to 1\,$\upmu$A even at $T=0.1$\,K in a different low-temperature setup (see Supp. Mat. Fig.~S10).
\newline
%
%%%%%%%%%%%%%%%%%%%%%%%%% Figure 4
\begin{figure}[t]
    \centering
    \includegraphics[width=0.9\linewidth]{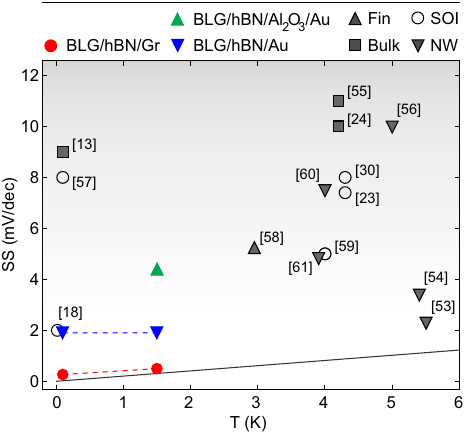}
    \caption{Comparison of the extracted low-temperature SS values for different types of FET devices. The red dots correspond to the device presented in this paper. The blue triangles refer to a similar device but where the top gate was made of gold instead of graphite, and the green triangle refers to a third BLG device with an additional \ch{Al2O3} layer between the hBN and the gold gate.  The empty symboles correspond to SS values reported in the literature for FETs based on different technologies (silicon on insulator (SOI), bulk CMOS, Fin, and nanowire FETs~\cite{Han2023, Han2022, Richtstein2022Oct, Incandela2018, Beckers2018Mar, Singh2006, Paz2020, Han2021, habicht2021, Sekiguchi2021, Beckers2020dec, Bohuslavskyi2019Mar, Galy2018may,paz2021}). FETs based on vdW heterostructures outperform all other technologies in terms of SS at cryogenic temperatures. The solid black line is the theoretical Boltzmann limit SS$_\text{BL}$ $= k_\text{B}T/e \ln(10)$.}
    \label{fig:f4}
\end{figure}
%%%%%%%%%%%%%%%%%%%%%%%%%
%
%
%
%
The measurements presented in Fig.~\ref{fig:f3} also show that the threshold voltage shifts to lower values of $V_\text{g}$ with increasing $V_\text{ds}$, without significantly affecting SS, see Fig.~\ref{fig:f3}a (more data are provided in Sec. 3 in the Supp. Mat.).
This implies that – despite the small on-current – the device presented in this manuscript could be operated at $T=0.1$\,K as a FET with an on-off ratio of at least 10$^5$ and an operational voltage range of only 3~-~4\,mV by suitably
choosing the drain-source voltage $V_\text{ds}$, thanks to the small SS $\approx250$\,$\upmu$V/dec. At $T=1.5$\,K, reaching an on-off ratio
of 10$^5$ will require operational voltages of 6~-~7\,mV due to a slightly
higher noise level and slightly higher SS $\approx500$\,$\upmu$V/dec.\newline
~\newline

Finally, we summarize in Fig.~\ref{fig:f4} the minimum inverse SS for different transistor device architectures reported in the literature (empty dots) as a function of temperature for low $T\leq6$\,K.
The best performing conventional FET devices, based on silicon-on-insulator~\cite{Galy2018may} or nanowires~\cite{Han2023}, allow to reach SS $\approx2$\,mV/dec. These values are almost an order of magnitude higher than the 250\,$\upmu$V/dec of the BLG-based devices reported in this work (red dots). 
The theoretical Boltzmann limit is included as a solid line. At $T=1.5$\,K, the Boltzmann limit is SS$_\text{BL}\approx 300$\,$\upmu$V/dec, only slightly less than the subthreshold slope of our device (SS $\approx 500$\,$\upmu$V/dec).  
We attribute this improvement in SS directly to the reduced interface disorder in devices based on pure vdW heterostructures, i.e., without bulk interfaces to metal or oxides.  The detrimental effect of bulk interfaces is well illustrated by the much higher SS values of BLG devices, where the top gate was made of gold instead of graphite (blue triangles in Fig.~\ref{fig:f4}). A BLG device with an additional \ch{Al2O3} between the metal top gate and the top hBN performed even worse (green triangle). \\
\newline
In summary, we have demonstrated that BLG devices based on pure vdW materials exhibit excellent band gap tunability and have provided evidence that 2D material-based FETs offer superior device behavior at cryogenic temperatures, with SS in the order of 250\,$\upmu$V/dec, only one order of magnitude above the Boltzmann limit of SS$_\text{BL}\approx20$\,$\upmu$V/dec at $T=0.1$\,K. 
The ability to also electrostatically confine carriers in BLG~\cite{Overweg2018Dec, Eich2018Aug, banszerus2023may} and the excellent performance as a field-effect transistor make this type of device an ideal platform for cryogenic applications
and calls for further device design improvements that allow for down-scaling and circuit integration. Moreover, we expect this work to trigger the exploration of pure vdW heterostructure FETs based on true 2D semiconductors, such as the transition metal dichalcogenides MoS$_2$ and WSe$_2$.

%\subsection*{Acknowledgements}
\textbf{Acknowledgements} The authors thank S.~Trellenkamp and F.~Lentz for their support in device fabrication. %
This project has received funding from the European Union's Horizon 2020 research and innovation programme under grant agreement No. 881603 (Graphene Flagship), from the European Research Council (ERC) under grant agreement No. 820254, the Deutsche Forschungsgemeinschaft (DFG, German Research Foundation) under Germany’s Excellence Strategy - Cluster of Excellence Matter and Light for Quantum Computing (ML4Q) EXC 2004/1 - 390534769, by the FLAG-ERA grant PhotoTBG, by the Deutsche Forschungsgemeinschaft (DFG, German Research Foundation) - 471733165, by the FLAG-ERA grant TATTOOS, by the Deutsche Forschungsgemeinschaft (DFG, German Research Foundation) – 437214324, from the EU project ATTOSWITCH under grant No. 101135571, and by the Helmholtz Nano Facility~\cite{Albrecht2017May}. K.W. and T.T. acknowledge support from the JSPS KAKENHI (Grant Numbers 20H00354, 21H05233 and 23H02052) and World Premier International Research Center Initiative (WPI), MEXT, Japan.

\textbf{Data availability}
The data supporting the findings are available in a
Zenodo repository under accession code 10.5281/zenodo.10526847.

%\clearpage

\bibliographystyle{unsrt}
\clearpage

%\newpage
%\renewcommand{\thefigure}{S\arabic{figure}}
%\setcounter{figure}{0}
%\section*{Supplementary Information}

\end{document}

\title{Supporting Information: \\Ultra-steep slope cryogenic FETs based on bilayer graphene}

\author{E.~Icking}
\affiliation{JARA-FIT and 2nd Institute of Physics, RWTH Aachen University, 52074 Aachen, Germany,~EU}%
\affiliation{Peter Gr\"unberg Institute  (PGI-9), Forschungszentrum J\"ulich, 52425 J\"ulich,~Germany,~EU}

\author{D.~Emmerich}
\affiliation{JARA-FIT and 2nd Institute of Physics, RWTH Aachen University, 52074 Aachen, Germany,~EU}%
\affiliation{Peter Gr\"unberg Institute  (PGI-9), Forschungszentrum J\"ulich, 52425 J\"ulich,~Germany,~EU}

\author{K.~Watanabe}
\affiliation{Research Center for Electronic and Optical Materials, National Institute for Materials Science, 1-1 Namiki, Tsukuba 305-0044, Japan
}
\author{T.~Taniguchi}
\affiliation{ 
Research Center for Materials Nanoarchitectonics, National Institute for Materials Science,  1-1 Namiki, Tsukuba 305-0044, Japan
}%

\author{B.~Beschoten}
\affiliation{JARA-FIT and 2nd Institute of Physics, RWTH Aachen University, 52074 Aachen, Germany,~EU}%
\author{M.~C.~Lemme}
\affiliation{Chair of Electronic Devices, RWTH Aachen University, 52074 Aachen, Germany,~EU}
\affiliation{AMO GmbH, 52074 Aachen, Germany,~EU}
\author{J.~Knoch}
\affiliation{IHT, RWTH Aachen University, 52074 Aachen, Germany,~EU}
\author{C.~Stampfer}
\affiliation{JARA-FIT and 2nd Institute of Physics, RWTH Aachen University, 52074 Aachen, Germany,~EU}%
\affiliation{Peter Gr\"unberg Institute  (PGI-9), Forschungszentrum J\"ulich, 52425 J\"ulich,~Germany,~EU}%

\date{\today}

\maketitle

In the first section, we present the equations used to calculate the band gap in bilayer graphene as a function of the displacement field based on a self-consistent approach following Ref.~\cite{McCann_2013} and Ref.~\cite{Slizovskiy2021}.
Furthermore, we provide additional information for the first sample introduced in Figs.~1 and 2 of the main manuscript: in Sec.~\ref{optical}, we present an optical image of the first device, in Sec.~\ref{QH}, we present quantum Hall measurements to estimate the gate-lever arms, in Sec.~\ref{DiamondSlopes} we discuss the slopes of the diamond outlines, in Sec.~\ref{Ig}, we explain how we extract the gate leakage current, and Sec.~\ref{1.5} presents some drain current traces for higher temperatures. As mentioned in the main text, additional information regarding the calculated band structure shown in Fig.~2d can be found in Sec.~\ref{calc}. In Sec.~\ref{2nd}, we present comparable data for a second device, similar to what we have shown in the main text for the first device. Finally, in Sec.~\ref{bulk}, we show the drain-current traces for the BLG devices with Au top gate and additional \ch{Al2O3} dielectric.

\section{Band gap as a function of displacement field}
The band gap in BLG~\cite{Min2007,McCann_2013} 
\begin{equation}
E_\mathrm{g}(\Delta)=\frac{|\Delta|}{\sqrt{1+(\Delta/\gamma_1)^2}},
\label{eq:gap}
\end{equation}
with the interlayer coupling strength $\gamma_1\approx 0.38$\,eV and the onsite potential difference~\cite{Kuzmenko2009Nov,Joucken2020,Jung2014Jan,McCann_206}
\begin{equation}
    \Delta=\frac{ d_0 e D}{\varepsilon_0\varepsilon_\text{BLG}}+\frac{d_0 e^2}{2\varepsilon_0 \varepsilon_\text{BLG}}\delta n(\Delta),
    \label{eq:Delta}
\end{equation}
depends on the strength of the applied displacement field $D$. Here, $d_0=0.34$\,nm is the interlayer spacing of BLG, and $\varepsilon_0$ is the vacuum permittivity~\cite{McCann_2013}. The effective dielectric constant of BLG $\varepsilon_\text{BLG}$ accounts for the electric susceptibility of each layer and the corresponding dielectric polarization~\cite{Slizovskiy2019Dec}.

\newpage
\section{Optical Image of the first device}\label{optical}

\begin{figure*}[hbt]
    \centering
    \includegraphics[width=0.4\linewidth]{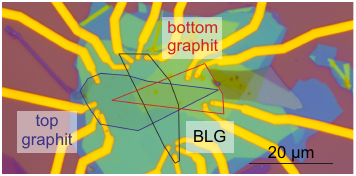}
    \caption[QH]{Optical image of the first device discussed in the main text. The BLG and the graphite gates are highlighted. The dual gated area is roughly $6 \times 9$\,$\upmu$m$^2$.}
    \label{fig:S_optical}
\end{figure*}
\FloatBarrier
\section{Quantum Hall measurements and gate lever-arm}\label{QH}

The gate-lever arm
\begin{equation}
    \alpha_\mathrm{tg,bg}=\frac{\nu e B}{h V_\mathrm{tg,bg}},
\end{equation}
can be extracted from quantum hall measurements by fitting the Landau level (see Fig.~\ref{QH})~\cite{Zhao2010,Dauber2017,SonntagMay2018,Schmitz2020}.
Here, $\nu$ is the Landau filling factor, $B$ the applied magnetic field, $h$ the Planck constant, and $V_\mathrm{tg(bg)}$ the applied gate voltage. A full list of the gate-lever arms, the relative gate lever arm (extracted from resistance maps as in Fig.~1c), and the thicknesses of the dielectrics can be found in Table~\ref{tab:lever arm} for both devices with graphite top and bottom gates.
\begin{figure*}[hbt]
    \centering
    \includegraphics[width=0.35\linewidth]{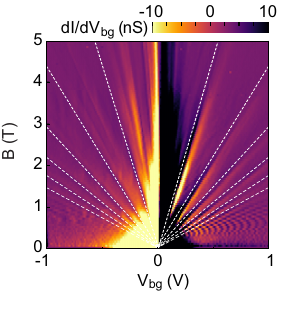}
    \caption[QH]{Transconductance as a function of out-of-plane magnetic field and bottom gate voltage. Fitting the Landau level (white dashed line) allows to extract the gate-lever arm.}
    \label{fig:S_f1}
\end{figure*}

\begin{table}[hbt]
    \centering
    \begin{tabular}{|c|c|c|c|c|}\hline
         Sample & $\alpha_\text{bg}(10^{11}$\,V$^{-1}$cm$^{-2})$ & $\beta$ & d$^\text{top}_\text{hBN}$ (nm) & d$^\text{bottom}_\text{hBN}$ (nm)\\\hline
         1 (main text) & 8.4 & 1.22 & 17 & 22\\\hline
         2 (Supp. Mat.) & 8.5 & 0.62 & 32 & 19 \\\hline
    \end{tabular}
    \caption{List of bottom gate lever arm extracted from QH measurements as depicted in Fig.~\ref{fig:S_f1}, relative lever arm from resistance maps (Fig.~1c), and hBN thicknesses from AFM measurements for both devices with a graphite top and bottom gate.}
    \label{tab:lever arm}
\end{table}

\FloatBarrier
\newpage

\section{Slopes of diamond outlines} \label{DiamondSlopes}
\begin{figure*}[hbt]
    \centering
    \includegraphics[width=0.5\linewidth]{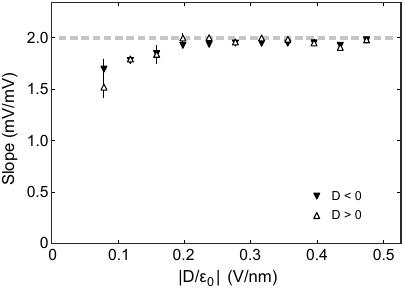}
    \caption[100mK]{The slopes of the outline of the diamonds (see black dashed lines in Fig.~1d in the main manuscript) obtained from finite bias spectroscopy measurements as a function of the applied displacement field $|D/\varepsilon_0|$. For details on the slope-extraction method, see Ref.~\cite{Icking2022Oct}.}
    \label{fig:S_DS}
\end{figure*}

The slope of the outline of the diamonds of suppressed conductance obtained from the finite bias spectroscopy measurement should have, in the ideal case, a slope of 2~\cite{Icking2022Oct}. A reduction in slope indicates that it is no longer possible to have a direct tuning of the electrochemical potential $\mu$ in the band gap, i.e., $\Delta V_\text{g} > \Delta \mu/e$. Interestingly, when plotting the extracted slope of the outline of the diamond (extracted by a fixed resistance threshold value $10^9\,\Omega$, exactly as also used in Ref.~\cite{Icking2022Oct}) as a function of the displacement field $D$, we observe a constant slope of very close to 2 for $|D/\varepsilon_0| \gtrsim 0.2$\,V/nm, see Fig.~\ref{fig:S_DS}.  
For smaller displacement fields ($D/\varepsilon_0 < 0.2$\,V/nm) our method of extracting the slope of the diamond outline breaks down, resulting in smaller values (as shown in Fig.~\ref{fig:S_DS}). This breakdown is mainly due to the rigidly fixed resistance threshold combined with the rather poor measurement resolution (see e.g. the leftmost panel in Fig.~1d in the main manuscript). In short, for such small displacement fields, the resolution of the diamonds is unfortunately insufficient to fully resolve any sharp structures that would allow reliable extraction of the slope of the diamond outline. As a result, the edge appears smeared, resulting in a decreasing slope value.

\FloatBarrier

\newpage
\section{Extracting the gate current} \label{Ig}

\begin{figure*}[hbt]
    \centering
    \includegraphics[width=1\linewidth]{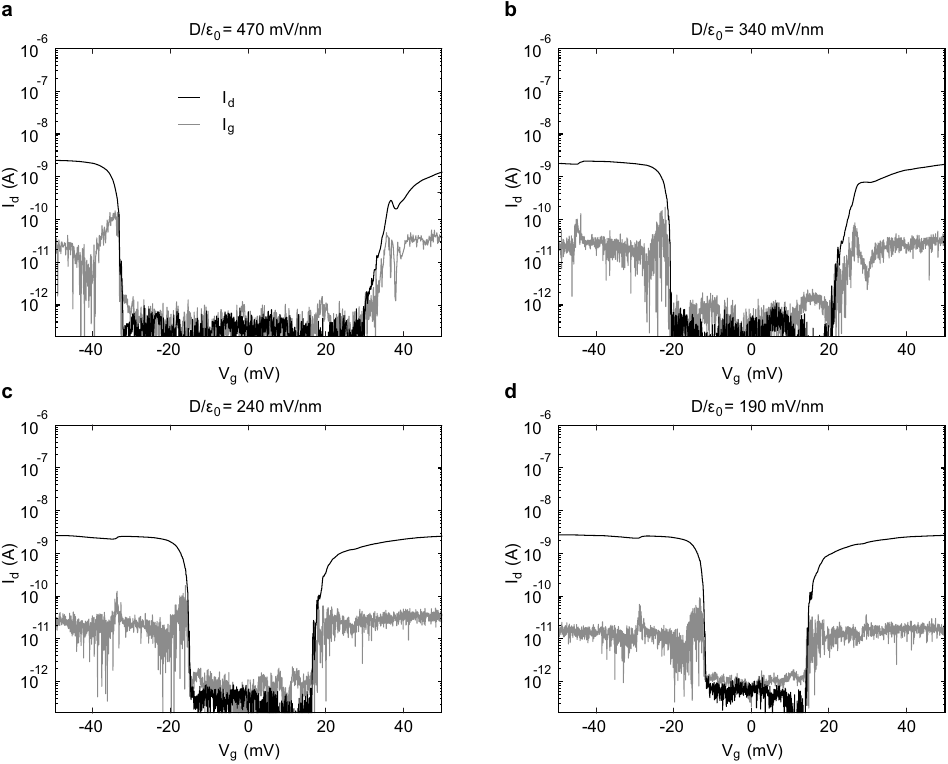}
    \caption[1500mK]{Drain-current (black) and gate current (gray) as a function of effective gating voltage $V_\text{g}$ at a temperature of $T=0.1$\,K at $V_\text{ds}=0.1$\,mV in the cases of the four different displacement fields as depicted in Fig.~2 in the main text.}
    \label{fig:S_Ig}
\end{figure*}

In the experiments, we apply the drain-source voltage symmetrically using an IV converter, which allows us to measure the source current $I_\text{s}$ and the drain current $I_\text{d}$ simultaneously. We can estimate the gate current $I_\text{g}\approx\Delta I$ from the difference $\Delta I = |I_\text{d}-I_\text{s}|$ between these two currents. Fig.~\ref{fig:S_Ig} shows the comparison between $I_\text{d}$ and $I_\text{g}$ for the four displacement fields, as shown in Fig.~2 in the main text.

\FloatBarrier

\newpage
\section{Subthreshold slope at 1.5~K} \label{1.5}

\begin{figure*}[hbt]
    \centering
    \includegraphics[width=1\linewidth]{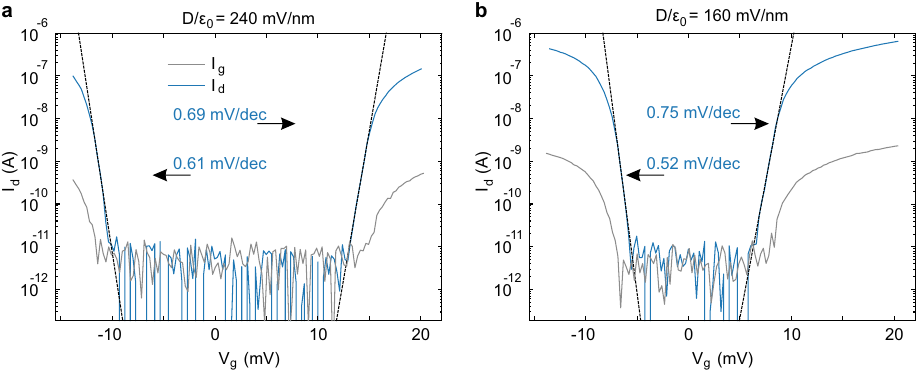}
    \caption[1500mK]{\textbf{a,b} Drain-current as a function of effective gating voltage $V_\text{g}$ at a temperature of $T=1.5$\,K obtained in a VTI cryostat at $V_\text{ds}=6$\,mV at $D/\varepsilon_0=0.24$\,V/nm (panel a) and 0.16\,V/nm (panel b). From the fit (black dashed line), the subthreshold slope for the two displacement fields can be extracted for the valence and conduction band edge (indicated by black arrows). The gate current for the same displacement fields is shown in gray.}
    \label{fig:S_f2}
\end{figure*}

In Fig.~4, data points are depicted obtained at $T=1.5$\,K for the first device in a pumped $^4$He cryostat (see main text). In Fig.~\ref{fig:S_f2}, we show full line traces for two different displacement fields. Even at $T=1.5$\,K, we still see a slight asymmetry in the extracted subthreshold slopes at conduction and valence band edge. 

\FloatBarrier
\section{Subthreshold Slope as a function of $V_\text{ds}$}\label{Vsd}

\begin{figure*}[hbt]
    \centering
    \includegraphics[width=0.45\linewidth]{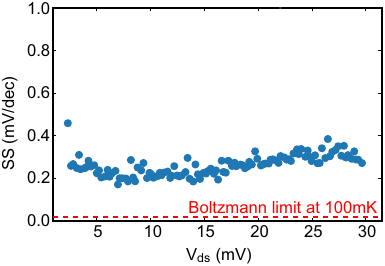}
    \caption[SS vs Vsd]{The extracted subthreshold slope for sample 1 for $D/\varepsilon_0=0.27$\,V/nm as a function of applied drain-source voltage $V_\text{ds}$ show an almost constant value of 250\,$\upmu$V/dec.}
    \label{fig:SS_vs_Vsd}
\end{figure*}

Fig.~3 in the main text depicts the drain-current for a fixed displacement field $D/\varepsilon_0=0.24$\,V/nm as a function of $V_\text{g}$ for three different $V_\text{ds}$. In order to verify that the SS is not affected by $V_\text{ds}$ we performed a more detailed analysis, see Fig.~\ref{fig:SS_vs_Vsd}. The first sample shows SS $\approx200$-$350$\,$\upmu$V/dec without any dependence on $V_\text{ds}$.

\FloatBarrier
\section{Bandstructure Calculations}\label{calc}
The bandstructure is calculated numerically using the Hamiltonian~\cite{McCann2013}
\begin{equation}
\label{hamiltonianK}
    H=\left(
    \begin{array}{c c c c} 
         \Delta/2& v_0\pi^\dagger & -v_4\pi^\dagger  &-v_3\pi\\
         v_0\pi& \Delta/2+\Delta' & \gamma_1  &-v_4\pi^\dagger\\
         -v_4\pi& \gamma_1 & -\Delta/2+\Delta'  & v_0\pi^\dagger\\
         -v_3\pi^\dagger& -v_4\pi & v_0\pi  & -\Delta/2\\
    \end{array}
    \right),
\end{equation}
with $\pi\equiv \hbar(\xi q_x+iq_y)$ and $\xi=\pm1$.
The parameter $v_i\equiv\frac{\sqrt{3}a}{2\hbar}\gamma_i$ is defined for each coupling parameter $\gamma_i$. These include the intralayer coupling $\gamma_0=3.16$~eV, the intralayer coupling $\gamma_1= 0.381$~eV between dimer sites on different layers, $\gamma_3=0.38$~eV accounts for coupling between non-dimer atoms on the two different layers, $\gamma_4=0.14$~eV for coupling between non-dimer atoms with dimer site atoms on the other layer, $\Delta'=0.015$~meV is the energy difference between dimer and non-dimer sites, and $\Delta$ the onsite-potential difference, responsible for the band gap formation~\cite{McCann2013, Jung2014Jan}.

\FloatBarrier
\section{Second device}\label{2nd}

\begin{figure*}[hbt]
    \centering
    \includegraphics[width=0.65\linewidth]{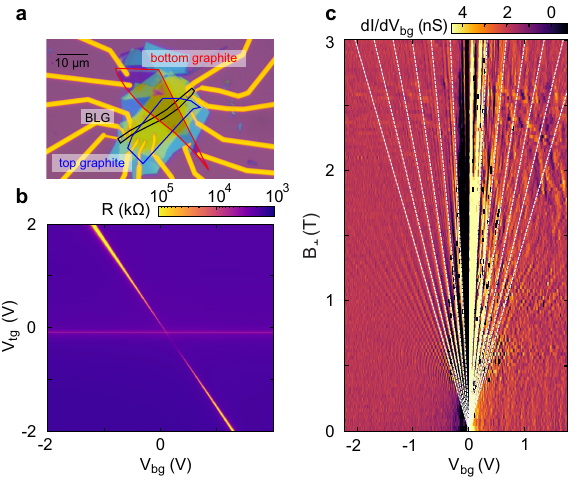}
    \caption[sample 2 diamonds]{\textbf{a} Optical image of a second device. The BLG and the graphite flakes are highlighted. \textbf{b} Resistance as a function of top and bottom gate voltage. The slope of the diagonal feature of increased resistance is equal to the relative lever-arm $\beta$. \textbf{c} Transconductance as a function of out-of-plane magnetic field and bottom gate voltage measured for the second device. }
    \label{fig:S_f3}
\end{figure*}

\begin{figure*}[hbt]
    \centering
    \includegraphics[width=0.5\linewidth]{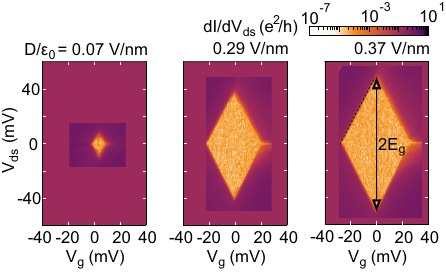}
    \caption[sample 2 diamonds]{Differential conductance as a function of effective gating voltage $V_\text{g}$ and drain-source voltage $V_\text{ds}$ for three different displacement fields measured in a second sample.}
    \label{fig:S_f4}
\end{figure*}

\begin{figure*}[hbt]
    \centering
    \includegraphics[width=0.3\linewidth]{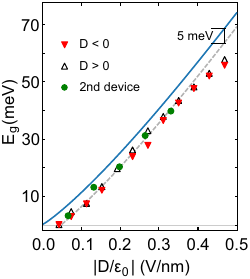}
    \caption[sample 2 gaps]{Extracted band gaps from finite bias spectroscopy measurements as a function of the displacement field. The results for the second device (green data points) are in good agreement with the result of the first device (triangles, see also main text). Both data sets follow the theory curve (blue line) if a small offset of about 5\,meV is included (dashed line).}
    \label{fig:S_f5}
\end{figure*}

\begin{figure*}[hbt]
    \centering
    \includegraphics[width=1\linewidth]{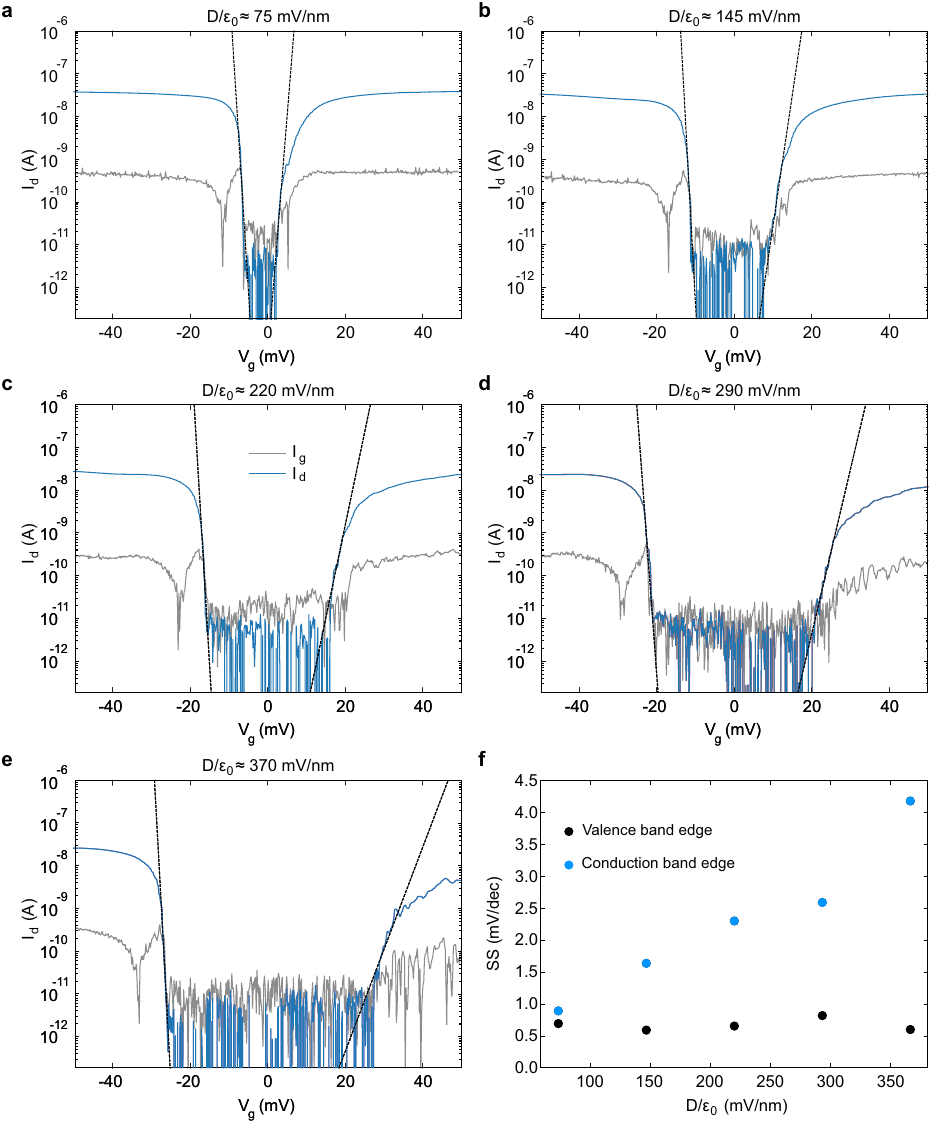}
    \caption[sample 2 subthreshold slopes]{\textbf{a} Drain current measured in a second device as a function of effective gating potential $V_\text{g}$ for five different displacement fields \textbf{a} $D/\varepsilon_0\approx0.07$\,V/nm, \textbf{b} 0.145\,V/nm, \textbf{c} 0.22\,V/nm, \textbf{d} 0.29\,V/nm, and \textbf{e} 0.37\,V/nm. \textbf{f} The fits (dashed lines in panels a-e) allow the extraction of the SS as a function of $D$ at the conductance and valence band edge. The same trend as in the main text for device 1 emerges: with increasing $D$, the SS extracted at the conductance band edge increases, while the SS extracted at the valence band edge stays nearly constant.}
    \label{fig:S_f6}
\end{figure*}

\begin{figure*}[hbt]
    \centering
    \includegraphics[width=0.45\linewidth]{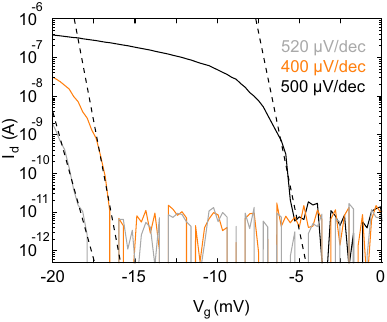}
    \caption[sample 2 SS vs Vsd]{Drain current as a function of effective gating potential $V_\text{g}$ at the valence band edge at $T=0.1$\,K measured in a dilution refrigerator for different applied drain-source voltages $V_\text{ds}=6$\,mV (gray), 10\,mV (orange) and 30\,mV (gray) at a fixed displacement field $D/\varepsilon_0\approx0.22$\,V/nm.}
    \label{fig:S_f7}
\end{figure*}

\begin{figure*}[hbt]
    \centering
    \includegraphics[width=0.45\linewidth]{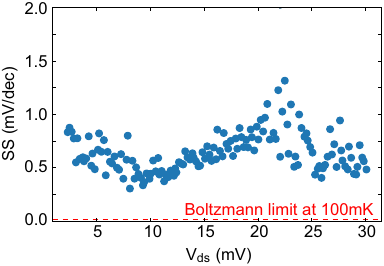}
    \caption[sample 2SS vs Vsd]{ The extracted SS of sample~2 for $D/\varepsilon_0=0.27$\,V/nm shows a much broader spread than the extracted SS values of sample~1. The extracted values have no clear dependence on the applied $V_\text{ds}$ and are more affected by sample-to-sample variations.}
    \label{fig:SS_vs_Vsd2}
\end{figure*}

Here, we introduce a second device (see an optical image in Fig.~\ref{fig:S_f3}a). This section shows the analog measurement for the measurements shown in Fig~1 and 2. From the resistance measurements (Fig.~\ref{fig:S_f3}b), we extract the relative gate lever-arm  (see Tab.~\ref{tab:lever arm}), from the quantum hall measurements (Fig.~\ref{fig:S_f3}c), we extract the bottom gate lever-arm (see Tab.~\ref{tab:lever arm}).\\
The finite bias spectroscopy (see Fig.~\ref{fig:S_f4}) also exhibits clean and highly symmetric diamonds. The extracted band gaps (see Fig.~\ref{fig:S_f5}) are in excellent agreement with the results from the main text.\\
We further measure the drain current for different displacement fields, similar to the analysis performed in Fig.~2 in the main text for the first device (see Fig.~\ref{fig:S_f6}a-e). The extracted subthreshold slopes (see Fig.~\ref{fig:S_f6}f) also show an asymmetry for values extracted at the valence and the conduction band edge, which increases with increasing displacement field. This effect is even more pronounced for the second device as we already have a significant difference of roughly 1\,mV/dec at $D/\varepsilon_0\approx150$\,mV/nm. The first device shows a comparable difference at slightly higher displacement fields $D/\varepsilon_0\geq250$\,mV/nm. Still, the overall trend is the same for both devices.\\
Measuring the drain current as a function of $V_\text{g}$ for different $V_\text{ds}$ exhibits higher on-currents of almost 1\,$\upmu$A, see Fig.~\ref{fig:S_f7}. In Fig.~\ref{fig:SS_vs_Vsd2}, we show the extended analysis of SS as a function of $V_\text{ds}$ for the second device, which yields SS values between 250\,$\upmu$V/dec up to 1\,mV/dec, without any significant dependency on $V_\text{ds}$. However, the spread in SS is slightly higher compared to the first device.

\begin{figure*}[hbt]
    \centering
    \includegraphics[width=0.55\linewidth]{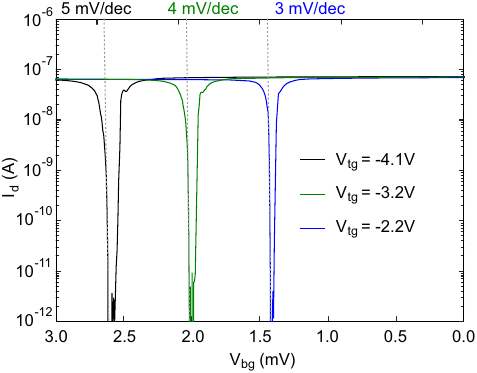}
    \caption[sample 2 tg operation]{Drain current as a function of bottom gate voltage, while the top gate is fixed to three different voltages. The resulting SS values are considerably worse compared to values extracted from measurements with fixed displacement fields.}
    \label{fig:S_f8}
\end{figure*}

\FloatBarrier
\section{Drain current in BLG devices with bulk interfaces}~\label{bulk}

\begin{figure*}[hbt]
    \centering
    \includegraphics[width=0.5\linewidth]{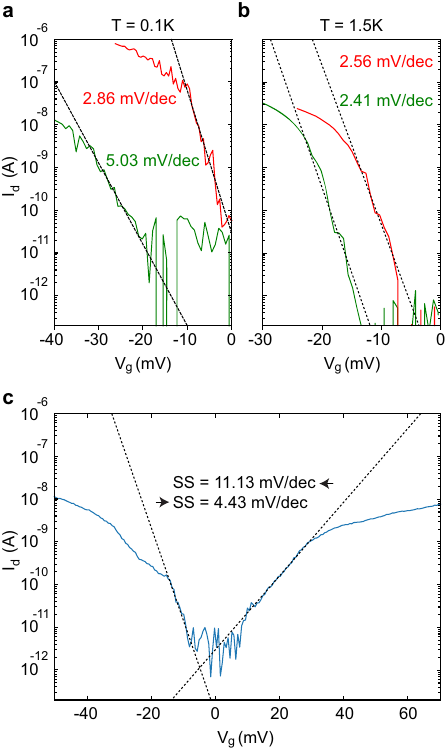}
    \caption[devices with additional bulk interfaces]{\textbf{a} Drain current as a function of effective gating voltage $V_\text{g}$ for two displacement fields $D/\varepsilon_0\approx0.19$\,V/nm (red) and 0.55\,V/nm (green) for a device with an Au top gate, measured at $T=1.5$\,K. \textbf{b} Drain current for two displacement fields $D/\varepsilon_0\approx0.33$\,V/nm (red) and 0.42\,V/nm (green) for a second device with an Au top gate, measured at $T=0.1$\,K. \textbf{c} Drain current measured in a device with an Au top gate and a layer of \ch{Al2O3} as an extra top gate dielectric (in addition to the top layer of hBN) at $D/\varepsilon_0\approx0.6$\,V/nm and $T=1.5$\,K.}
    \label{fig:S_f9}
\end{figure*}

In Fig.~4 of the main text, we present subthreshold slope values for two Gr/hBN/BLG/hBN/Au and one \newline Gr/hBN/BLG/hBN/\ch{Al2O3} device. Fig.~\ref{fig:S_f9}a shows the drain current as a function of effective gate voltage for the first device with an Au top gate measured at $T=0.1$\,K,  Fig.~\ref{fig:S_f9}b for the second device with an Au top gate measured at $T=1.5$\,K, and Fig.~\ref{fig:S_f9}c for the Gr/hBN/BLG/hBN/\ch{Al2O3}/Au device measured also at $T=1.5$\,K.

\bibliographystyle{unsrt}